\documentclass[doublecol]{epl2} 
\usepackage{graphicx}
\usepackage{amssymb}
\usepackage{graphicx}
\usepackage{amsmath,amssymb}
\usepackage[normalem]{ulem} 

\newcommand{\calG}{\mathcal{G}}

\title{Protein unfolding and refolding as transitions through virtual states}

\author{L.\ L.\ Bonilla$^\ast$, A.\ Carpio$^\dagger$, and A.\ Prados$^\ddagger$}
\shortauthor{L.L. Bonilla \etal}

\institute{$^\ast$G. Mill\'an Institute, Fluid Dynamics, Nanoscience and Industrial
Mathematics, 
Universidad Carlos III de Madrid, 28911 Legan\'es, 
Spain; $^\dagger$Departamento de Matem\'atica Aplicada, 
Universidad Complutense de Madrid, 28040 Madrid, 
Spain; $^\ddagger$F\'{\i}sica Te\'{o}rica, 
Universidad de Sevilla,
Apartado de Correos 1065, E-41080, Sevilla, 
Spain, EU
}

{\pacs{87.15.Cc}{ Folding: thermodynamics, statistical mechanics, models, and pathways}}
{\pacs{05.40.-a}{ Fluctuation phenomena, random processes, noise, and Brownian motion}}
{\pacs{87.14.et}{ Generic models (lattice, HP, etc.)}}

\abstract{Single-molecule atomic force spectroscopy probes elastic
  properties of titin, ubiquitin and other relevant proteins. We
  explain bioprotein folding dynamics under both length- and
  force-clamp by modeling polyprotein modules as particles in a
  bistable potential, weakly connected by harmonic spring
  linkers. Multistability of equilibrium extensions provides the
  characteristic sawtooth force-extension curve. We show that abrupt
  or stepwise unfolding and refolding under force-clamp conditions
  involve transitions through virtual states (which are
  quasi-stationary domain configurations) modified by thermal
  noise. These predictions agree with experimental observations.}

\begin{document}

\maketitle 

\section{Introduction}

The study of single molecules may explain the function of many
molecular assemblies found in cells
\cite{alb98,obe08,lin08,thirumalai13}. Tissue elasticity in living
organisms results from the extension and recoil of proteins fastened
to rigid structures that move under force. Polyproteins or modular
proteins, such as titin that plays an important role in muscle
contraction \cite{lin08}, ubiquitin and other relevant proteins
\cite{car99,fis00,MyD12,rit06jpcm}, comprise a number of repeated
single protein domains joined by short peptide linkers. To reduce the
variety of single protein domains, artificial homopolyproteins
comprising a number of identical protein modules (thereby having the
same mechanical properties) have been engineered by using molecular
biology techniques \cite{car99}.

A simple version of tissue elasticity appears in most single-molecule
experiments, like atomic force microscopy (AFM), in which a
biomolecule is chained to rigid platforms whose motion is controlled
\cite{rit06jpcm}. As the polyprotein is pulled, one or more modules
unfold at a typical force that measures its mechanical stability.  It
should be stressed that the unraveling of a domain is a stochastic
event and may occur in a certain range of forces. These
\textrm{length-controlled} experiments deliver a sawtooth force-extension
curve (FEC) \cite{fis00,car99,lip01,bus03,lip02,PCyB13}.  Similar
curves are obtained by stretching nucleic acids
\cite{lip01,smi96,bus03,hug10} and other biomolecules
\cite{fis00,cao08}. When the FEC is swept at a finite rate, stochastic
jumps between folded and unfolded states may be observed
\cite{lip01,lip02,hug10,PCyB12}, and the unfolding force increases
with the extension rate.

In a typical force-clamp experiment, the force is first raised, kept
at a large value until all domains become unfolded and then abruptly
lowered to a smaller value \cite{li04,ber10}. Immediately after the
force increment, abrupt or stepwise unfolding of the polyprotein
follows \cite{wal07,lan12}. On the other hand, after the force is
lowered, refolding is similar for single module proteins \cite{ber10}
and for homopolyproteins \cite{li04}; the folding events do not show
traces of sequential folding for polyproteins (see Figure S2 of
Ref.~\cite{li04} for the only reported exception showing stepwise
folding of a two-module protein). Berkovich et al. \cite{ber10}
interpret the results of their single-module protein experiments using
a simple Langevin equation model that includes an effective bistable
potential for a range of the applied force.

The sawtooth FEC is well understood: force jumps are already present
in equilibrium \textrm{when length is controlled
\cite{PCyB13,ByG11,tommasi}}.  However, \textrm{force-clamp
  experiments are not: S}light changes of the forces lead to
completely different behaviors, \textrm{suggesting} that polyproteins are
operating near critical conditions \textrm{therein}. We
\textrm{now put forward and motivate}  a simple model able to explain some aspects of polyprotein
folding and refolding under \textrm{either length or force
  control}. \textrm{It} is inspired in mathematically similar spatially
discrete models for charge transport in weakly coupled semiconductor
superlattices (SLs) \cite{BGr05}. SLs also have a sawtooth
current-voltage curve (similar to FEC in polyproteins) under voltage
bias (voltage is analogous to extension, current to
force). \textrm{A related model for shape-memory alloys \cite{PyT02} has been
  recently reworked to analyze the FEC of biomolecules
  \cite{ByG11,tommasi}. Moreover, behavior resembling stepwise unfolding is observed in overdamped  Frenkel-Kontorova (FK) chains (which have bistable on-site potentials and nearest neighbor harmonic coupling) \cite{car01,car03} and in chains with a bistable nearest neighbor snap-spring
  potential that become the FK model \cite{TyV10}.}

\begin{figure}[htbp]
\begin{center}
\includegraphics[width=3in]{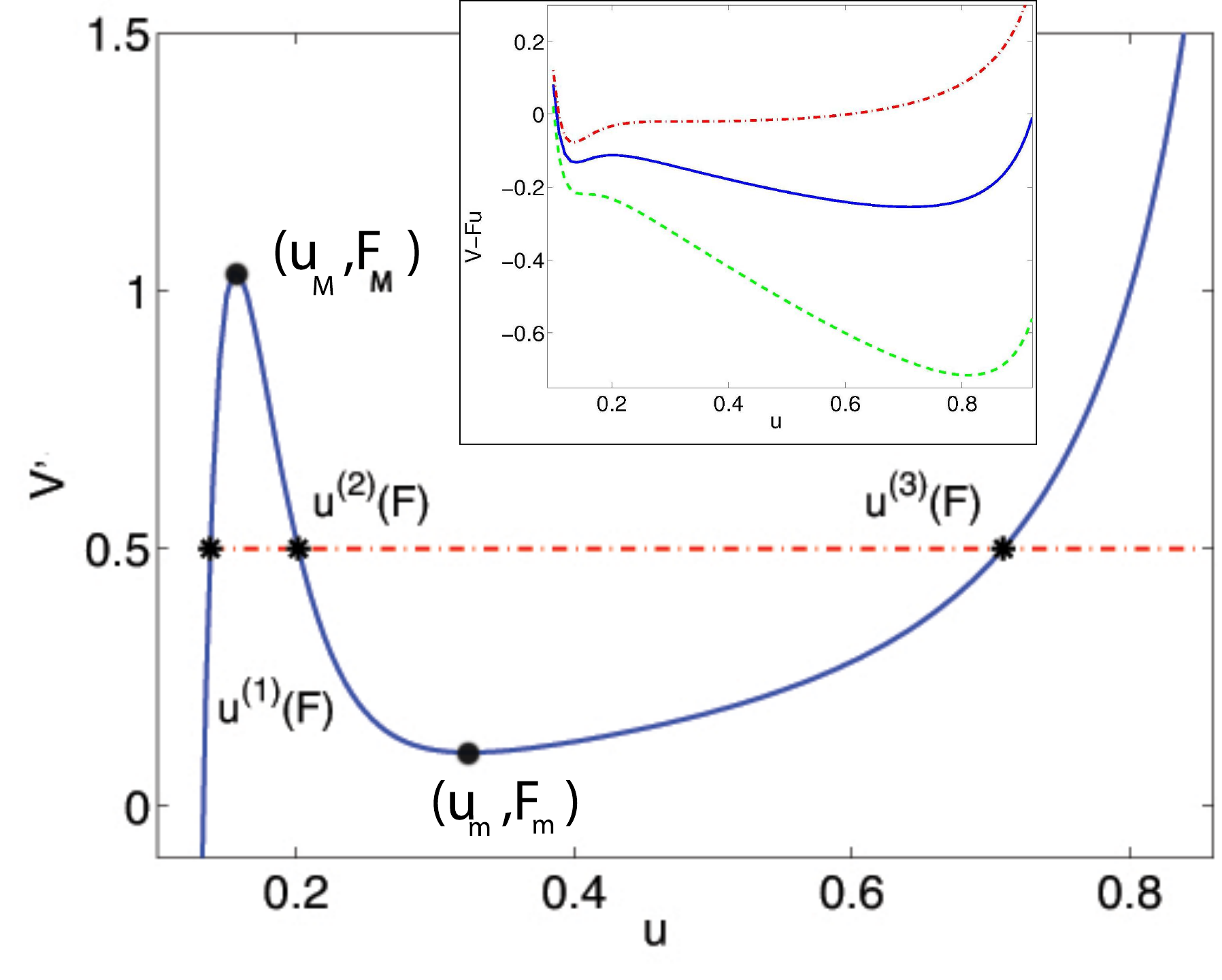}
\caption{Force field $V'(u)$ vs.\/ $u$ and the three solutions of
$V'(u)=F$, $u^{(1)}(F)<u^{(2)}(F)<u^{(3)}(F)$ in the force range
$F_m<F<F_M$. Therein, $F_M$ (about $104$pN or $1.04$ in dimensionless
variables) and $F_m$ (about $10.4$pN or $0.104$) are the local maximum
and minimum forces, with corresponding extensions $u_M$ and $u_m$,
respectively. The unit of force is $[F]=100$pN, and $L_c=30$nm.
Inset: Potential $V(u)-Fu$ of ubiquitin, at $F=10$, $50$ and $100$pN
(from top to bottom).}
\label{fig2a}
\end{center}
\end{figure}

\section{\textrm{Model}}
The time scale for stress relaxation inside a module is much smaller
than the time scale of a typical unfolding/refolding event and
therefore we can assume instantaneous mechanical equilibrium inside
each module at the time scale of unfolding/refolding events. \textrm{Then each} module of
extension $u$ \textrm{is modeled} as a particle
in a bistable potential whose minima represent folded (enthalpic
minimum) and unfolded (entropic minimum) states  \textrm{\cite{ber10}}, see
Fig~\ref{fig2a}. \textrm{T}he \textrm{following} effective potential \textrm{is shown to provide a good description of single-module proteins at temperature $T$ and zero external force \cite{ber10}:} 
\begin{eqnarray}
V(u) & = & U_0\!\left[\!\left(1-e^{-2b(u-R_c)/R_c}\right)^2-1\right]\! \nonumber\\&&
+\frac{k_BTL_c}{4P}\!\left(\frac{1}{1-\frac{u}{L_c}}-1-\frac{u}{L_c}
+\frac{2u^2}{L_c^2}\right)\!,   \label{a1}
\end{eqnarray}
For ubiquitin the applied force ranges from $10$ to $120$pN,
$P=0.28$nm (persistence length), $L_c=30$nm, $U_0=200$pN nm($\sim\!\!
48 k_B T$), $R_c=4$nm, $b=2$, $T=300$K.  In AFM experiments, the
polyprotein is tethered to two platforms and stretched so its geometry
is quasi-one-dimensional. If forces $\pm F$ are applied to the ends of
the modular protein and the $j$th module extends from $x_j$ to
$x_{j+1}$, with $u_j=x_{j+1}-x_j$, the potential energy due to the
force is $Fx_1-Fx_{N+1}=-F\sum_{j=1}^N u_j$. As part of the tertiary
structure of the polyprotein, modules are weakly interconnected in a
structure-dependent way. This weak interaction acts on the
unfolding/refolding time scale and tries to bring the extensions of
the modules to a common value corresponding to global mechanical
equilibrium. This crucial feature to explain sequential unfolding is
absent in simpler models that do not assign different elongations to
different modules \cite{ber10}.  As a simplification, we assume that
neighboring modules $(j-1,j)$ interact via a spring potential
$\frac{k}{2}(u_j-u_{j-1})^2=\frac{k}{2}(x_{j+1}-2x_j+x_{j-1})^2$,
where $k=0.0566 [F]/L_c$ is the spring constant.

The modules satisfy overdamped Langevin equations:
\begin{eqnarray}
 \gamma\dot{u}_j = -\frac{\partial}{\partial u_j} \calG(\bm{u},F,T)+ \sqrt{2D}\,\gamma\, \xi_j(t),  \label{a2}
\end{eqnarray}
where $\calG(\mathbf{u},F,T)=\sum_{j=1}^N\!\left[V(u_j)-Fu_j
+\frac{k}{2}(u_j-u_{j-1})^2 \right]$ is the overall potential for a
$N$-module protein, and $D=k_B T/\gamma=1000$ nm$^2$/s, $\xi_j(t)$,
$k_B$, and $T$ are the diffusion coefficient, zero-mean
delta-correlated independent identically-distributed white noises, the
Boltzmann constant and the temperature, respectively. Assuming
infinitely rigid springs connect the protein to AFM cantilever and
platform, $u_0=u_1$, $u_{N+1} =u_N$.
\begin{figure}
\begin{center}
\includegraphics[width=3in]{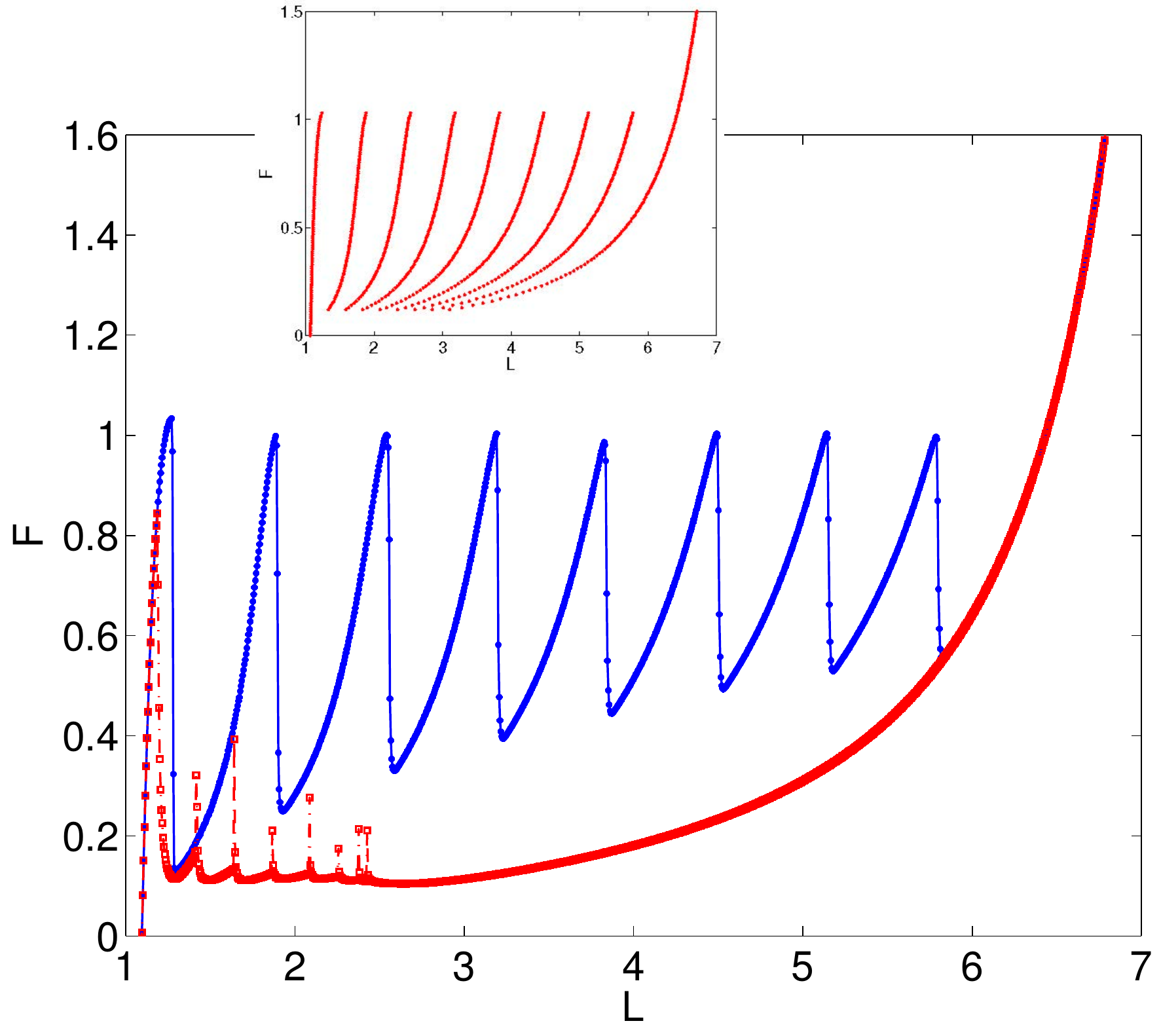}
\caption{Force-extension curve obtained by first solving numerically the
zero-noise Langevin equations with sudden length increases $\Delta
L=0.005$ applied at times $t_j=j\Delta t$, with $\Delta t=0.25$, until
all modules are unfolded at time $t_M$. After this time, we reverse
this procedure by applying length decrements $-\Delta L$ at times
$t_{M+j}= t_M+j\Delta t$ until we return to the initial state having
all modules folded. Time unit: $[t]=\gamma L_c/[F]=1.24$ms.  Inset:
Force vs length curves for the stationary solutions of the zero-noise
Langevin equations with 8 modules.
 }
\label{fig2b}
\end{center}
\end{figure}

\section{Deterministic dynamics}
The stochastic nature of the unfolding/refolding events is well
documented by experiments. However, the mere existence of stepwise
unfolding indicates that a deterministic scaffolding lies below such
events. To understand them, we first reveal the reason for their
existence by studying the dynamics of our system without noise and
later explain the important modifications noise brings to the picture.

In absence of noise and with time-independent length or force, the
system (\ref{a2}) has stable stationary configurations of folded and
unfolded modules with respective extensions $u^{(1)}(F)$ and
$u^{(3)}(F)$ in the metastability region $F_m <F< F_M$, defined in
Figure \ref{fig2a}. Configurations with only one \textit{domain wall},
separating domains of folded and unfolded modules, are stationary
(pinned) wave fronts. For a given number of unfolded modules, these
pinned wave fronts minimize the linkers contribution to energy and are
therefore the most stable configurations. By slowly increasing the
protein length with time and decreasing it after all modules unfold,
there appears the sawtooth FEC in the main panel. The system moves
over the stationary branches (as many as polyprotein modules) in the
inset of Fig.~\ref{fig2b}. In the pulling (resp.\/ pushing) process,
the system sweeps the branch where it was when the force variation
started until it reaches the limit of stability $F_M$ (resp.\/ $F_m$),
and then jumps to the next branch having one more (resp.\/ less)
unfolded module. The small upward jumps (refolding events) in the
pushing process have been observed in experiments, see Figs.~1C and 1D
of Ref.~\cite{lee10}. Thermal noise introduces fluctuations in this
folding/refolding diagram and changes the maxima and minima of Fig.\
\ref{fig2b}. In general, the FEC lies between the \textit{adiabatic
  limit at zero temperature} in the main panel of Fig.~\ref{fig2b},
and the \textit{quasistatic limit} discussed in Ref.~\cite{PCyB13}.

\begin{figure}[htbp]
\begin{center}
\includegraphics[width=3in]{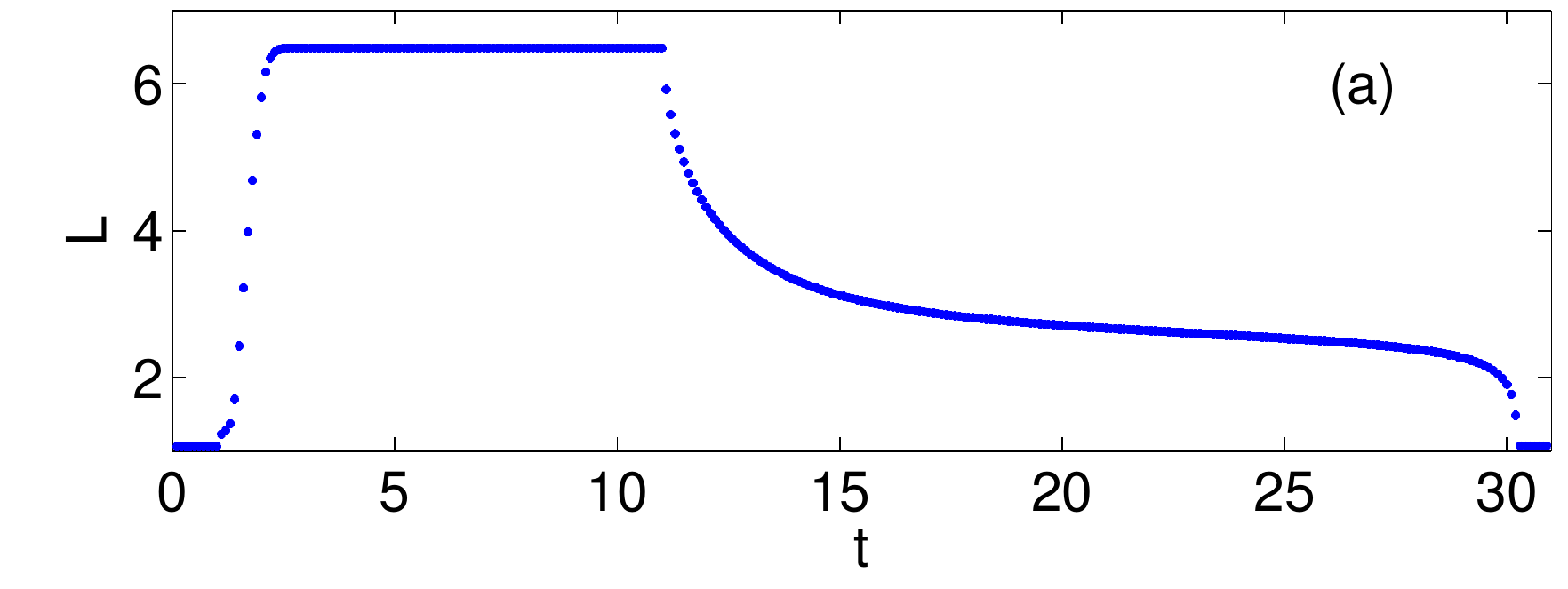}
\includegraphics[width=3in]{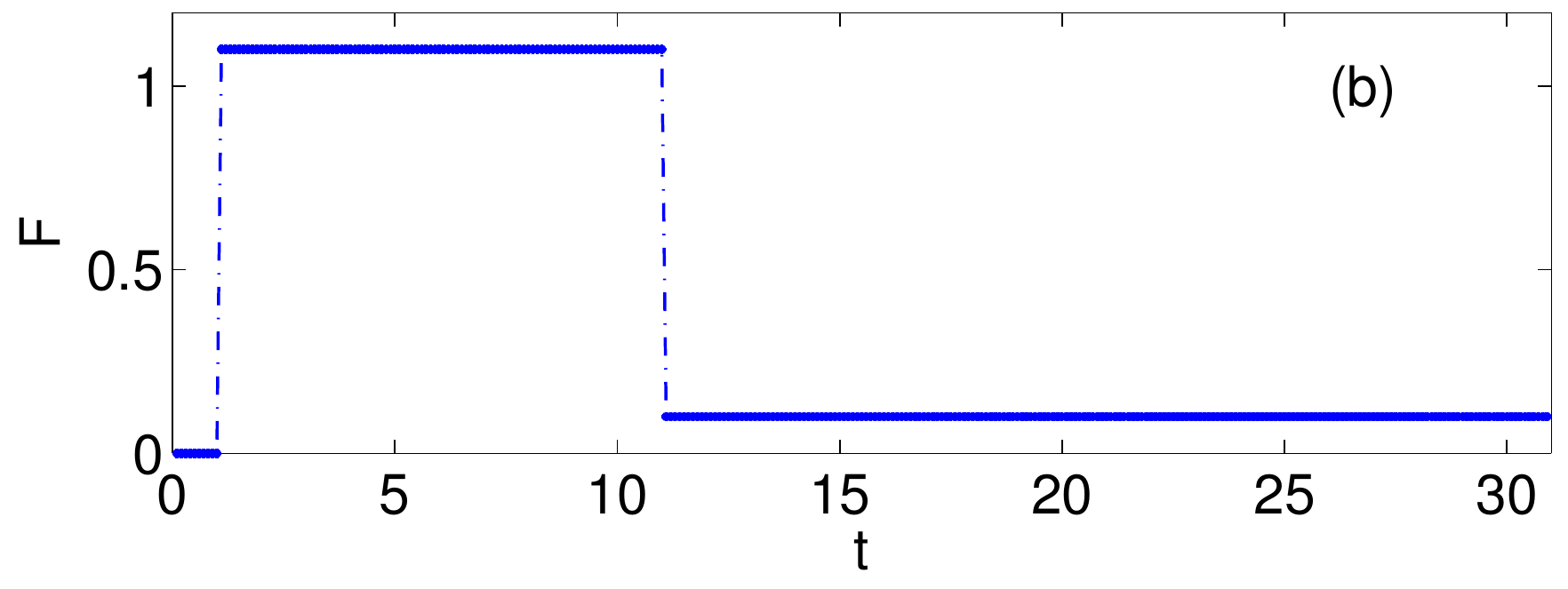}
\begin{tabular}{p{1.5in} p{1.5in}}
  \vspace{0pt} \includegraphics[width=1.5in]{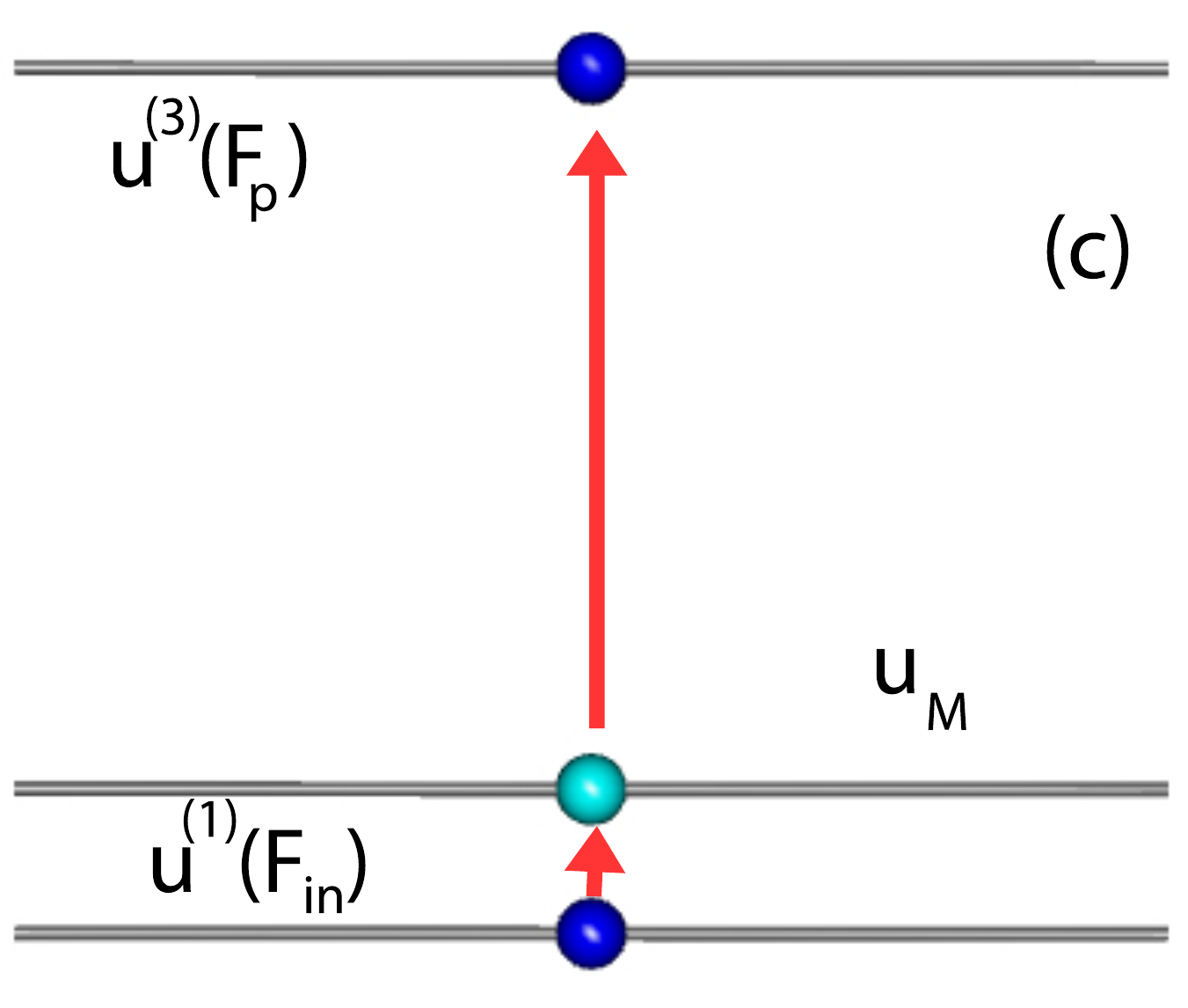} &
  \vspace{0pt} \includegraphics[width=1.5in]{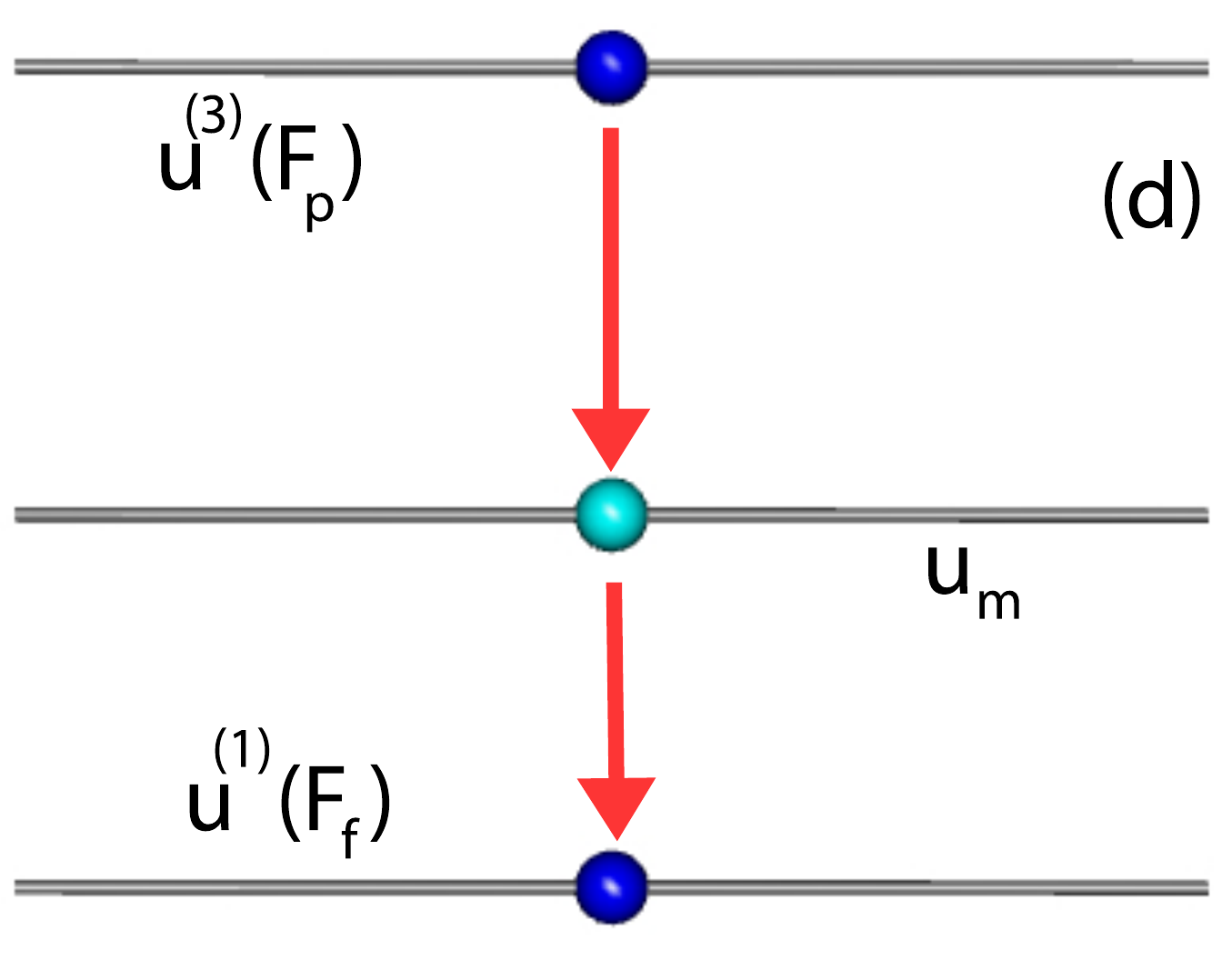}
\end{tabular}
\end{center}
\caption{(a) Protein length response to the force variation shown in
(b). Initially $F_{in}=0$ and $u_j=u^{(1)}(0)$ for all $j$. Peak
$F_p=1.1$ (110 pN) and final $F_f=0.1$ (10 pN) forces are slightly
larger than $F_M$ and slightly below $F_m$, respectively. (c) Protein
unfolding after $F_{in}\to F_p$: all the modules stretch to the
virtual state with extension $u_M$, spend some time there, 
then simultaneously unfold to reach extensions
$u^{(3)}(F_p)$. (d) Refolding stage: modules simultaneously
contract to $u_m$, stay there a long time
$\sim\pi\sqrt{2}/\sqrt{(F_m-F_f)V'''(u_m)}$, and refold to 
$u^{(1)}(F_f)$.  The virtual state has a much more noticeable effect
on the length response curve during refolding: $u_M-u^{(1)}(F)$
is small for all $F$ while $u^{(3)}(F)-u_m$ is not, see
Fig.~\ref{fig2a}.}
\label{fig3}
\end{figure}

\begin{figure}[htbp]
\begin{center}
\includegraphics[width=3in]{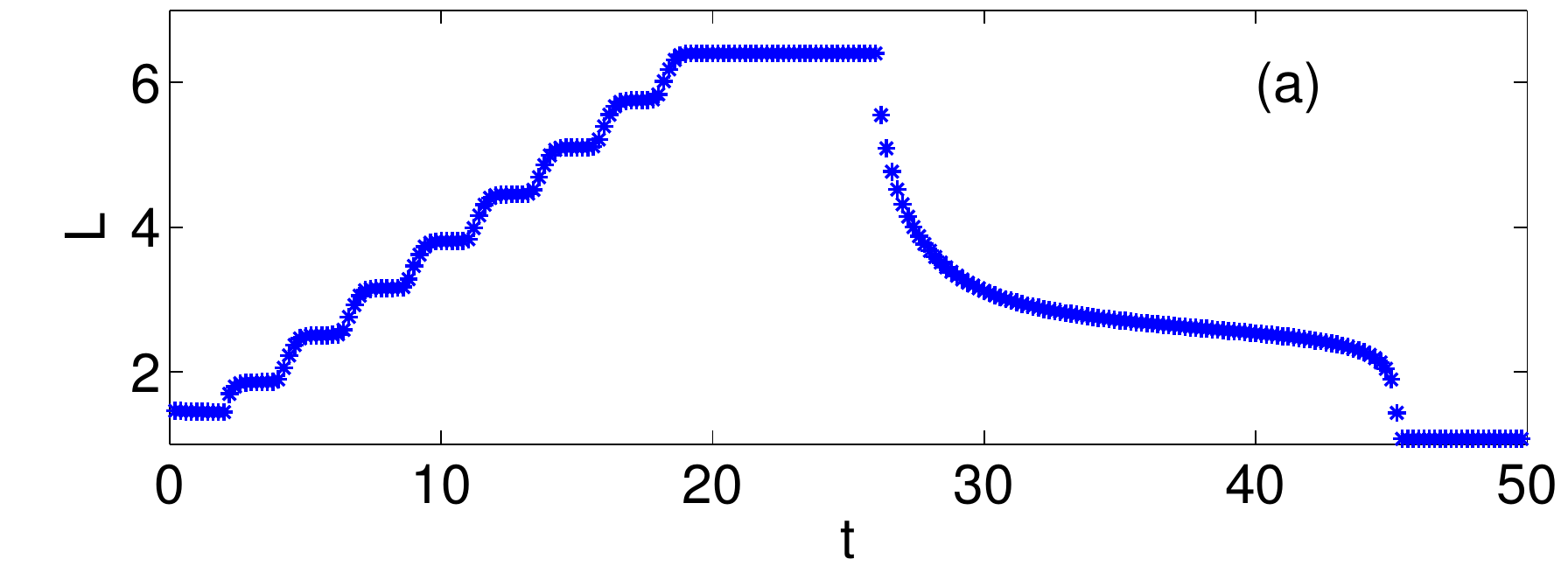}
\includegraphics[width=3in]{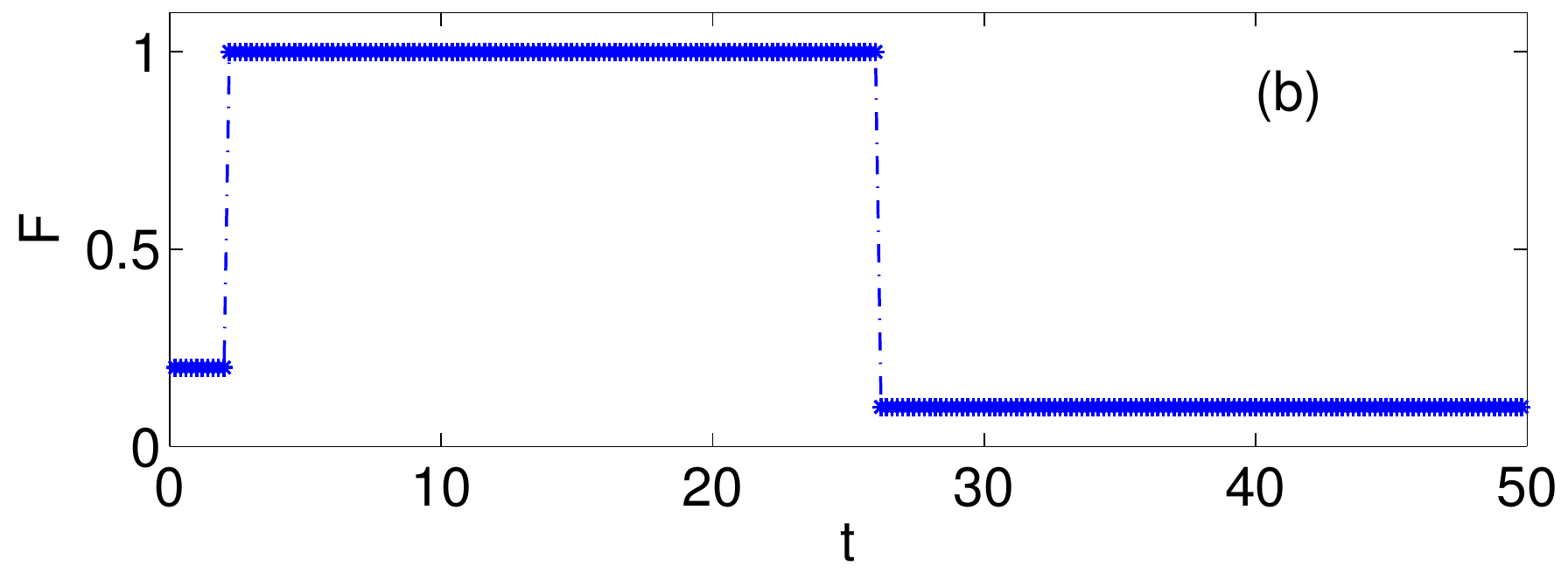}
\includegraphics[width=1.2in]{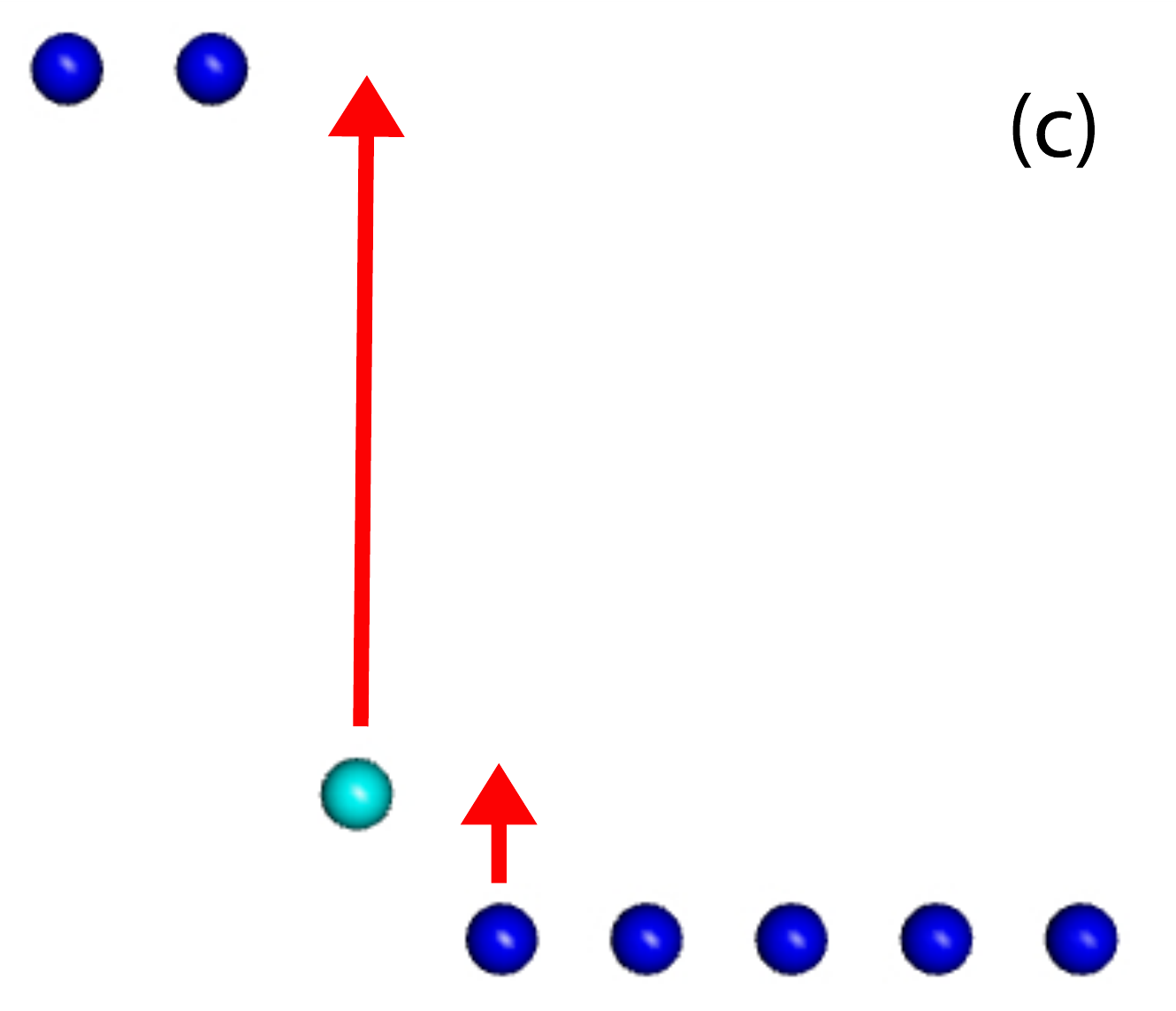}
\includegraphics[width=1.2in]{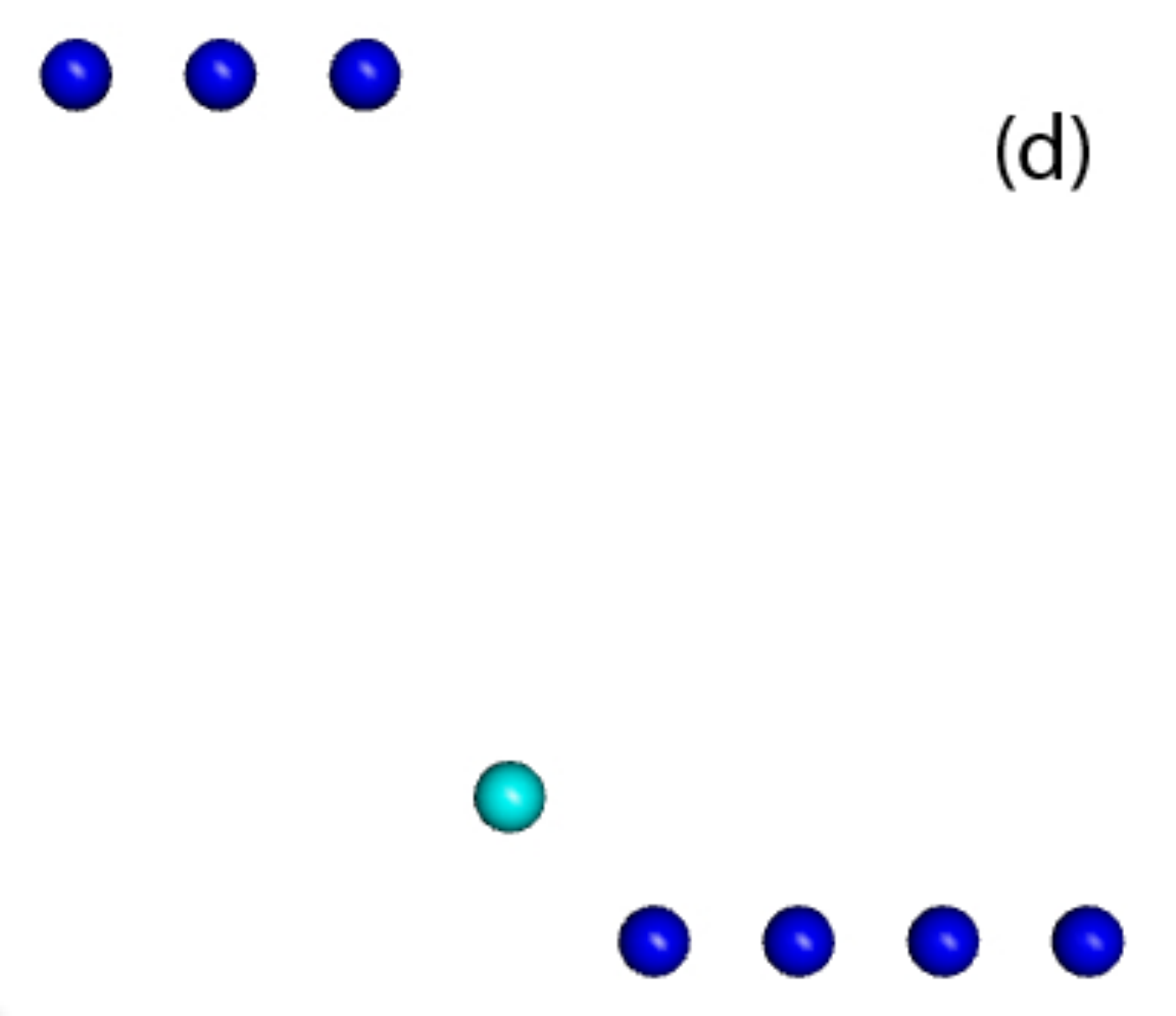}
\end{center}
\caption{
 (a) Protein length response to the force variation shown in (b) with
$F_{in}=0.2$, $F_p=1$ such that $F_{c2}<F_p<F_M$, $F_f=0.1$ (20, 100
and 10 pN, respectively). Initial extensions:
$u_j=u^{(1)}(F_{\text{in}})$ for $j>1$,
$u_1=u^{(3)}(F_{\text{in}})$. (c), (d) Sketches of the unfolding of
the active module at the domain wall, which here moves to the right,
leading to a new quasi-stationary configuration with one more unfolded
module. The long common time between jumps is
$\pi\sqrt{2}/\sqrt{(F_p-F_{c2})|V'''(u^*)|}$.
}
\label{fig4}
\end{figure}

In force-clamp experiments, the force first suddenly increases from
$F|_{t=0}\equiv F_{\text{in}}$ to a peak value $F_p$, remains there
for a given time, then abruptly decreases to $F_f$. Depending on
$F_p$, the polyprotein length increases either abruptly (large $F_p$,
as in Fig.~\ref{fig3}(a-b)) or in a succession of length jumps
(smaller $F_p$, as in Fig.~\ref{fig4}(a-b)). Depending on $F_f$, 
modules may simultaneously refold or remain unfolded at a
smaller length. We will show that this behavior arises because 
$F_p$ and $F_f$ are close to the \textit{critical}
forces $F_M$ and $F_m$, respectively. \textit{Virtual} states with
extensions $u_M$ or $u_m$ shown in Fig.~\ref{fig2a} play a crucial
role.

To understand the sudden unfolding in Fig.~\ref{fig3}(a), we assume
that initially all modules are equally folded, take $F_p$ just above
the local maximum $F_M$ and $F_f$ just below the local minimum $F_m$,
see Fig.~\ref{fig3}(b). Protein unfolding and refolding occur as
passages through virtual states. For $F=F_p$, no stable folded state
exists. Thus all modules jump in a short time after the force increase
to $F_p$ to the virtual state with extension $u_M=u^{(1)}(F_{M})$. The
modules remain there for a time (larger the smaller $F_p-F_M$ is,
infinite if $F_p=F_M$), until all modules unfold simultaneously to
acquire extensions $u^{(3)}(F_p)$. A subsequent sudden decrease to
$F_f$ \textit{just below the local minimum of the force field} where
$u^{(2)}(F_m)=u^{(3)}(F_m)=u_m$ makes all modules collapse
simultaneously to the folded state in a three-stage sequence, as
observed in experiments. Similarly, all modules first fall to the
virtual (unfolded) state with extension $u_m= u^{(3)}(F_m)$, stay
there for a long time, then abruptly refold to $u^{(1)}(F_f)$, see
Fig.~\ref{fig4}(a).  Had we chosen $F_f>F_m$, all modules would have
remained in the stable unfolded state $u^{(3)}(F_f)$. They would have
folded simultaneously and rapidly to $u^{(1)}(F_f)$ had $F_f$ been
smaller than but not close to $F_m$. This two-stage behavior is
similar to experimental observations, (see Fig.~9 of \cite{gar07}). Of
course, thermal noise is present in experiments ($T\neq 0$), each
module may spontaneously change from folded to unfolded state and
back, and the system may end up either in the unfolded or the folded
state for different realizations of the experiment.  See the next
section for more details.

The stepwise length increase in Fig.~\ref{fig4}(a) stems from a more
complicated dynamical behavior, appearing for $F_p<F_M$ but close
thereto as in Fig.~\ref{fig4}(b)). Let us consider an embryonic wave
front configuration, see Figs.\ \ref{fig4}(c) and (d).  Such
stationary configurations become unstable and the resulting wave front
moves with nonzero velocity if the force is on any of the two
depinning intervals $(F_m,F_{c1})$ and $(F_{c2},F_M)$, whose widths
vanish proportionally to the elastic constant $k$ in Eq.~\eqref{a1};
see the appendix and also Refs.~\cite{car01,car03}. New virtual states
appear, those corresponding to the Peierls critical forces
for wave front depinning \cite{car01,car03,TyV10}, $F_{c1}$
and $F_{c2}$, and more involved transitions through them play a
crucial role. See \cite{fis98} for general depinning
  transitions in random media.

After $F$ abruptly increases to $F_p$ in the depinning interval
$F_{c2}<F_p<F_M$, the modules evolve to a \textit{virtual}
quasi-stationary state for $F=F_{c2}$, with only one unfolded module
of extension $u^{(3)}(F_{c2})$. The adjacent module has an extension
$u^*$, slightly larger than that of the others, $u^{(1)}(F_{c2})$, see
Figure \ref{fig4}(c). This is the \textit{active} module: it is the
only one whose extension changes noticeably, slowly increasing from
$u^*$ until, at a precise time, it suddenly unfolds to
$u^{(3)}(F_{c2})$. Simultaneously, the next module becomes active
attaining extension $u^*$, see Fig.~\ref{fig4}(d). This saltatory
motion of the wave front continues until all the modules unfold to
$u^{(3)}(F_{c2})$, with all the time steps having the same length. In
the next section, we see that thermal noise makes the steps have
different lengths, as observed in the experiments, see for instance
Fig.~10 of \cite{gar07}.

An initial embryonic wave front configuration may be attained in two
ways: (i) For $F_{in}$ between $F_m$ and $F_p<F_M$, we put the system
in a configuration with only one unfolded module, that is, a point
close to the bottom of the second branch in Fig.~\ref{fig2b}; (ii) all
the modules are folded, but one of them has a slightly larger protein
length, e.~g.~ the one attached to the AFM {cantilever}. Accordingly,
its potential is $V(\mu_1u_1)$, $\mu_1<1$, and the corresponding local
maximum of the force field occurs at $\mu_1 F_M<F_M$ with a larger
extension $u_M/\mu_1$. If $\mu_1F_M<F_p<F_M$, this module unfolds
first, creates the wave front, and sequential stepwise unfolding
follows.

\section{Stochastic dynamics}

Considering white noise forces, the threshold
forces change because the modules may unfold (refold) for peak (final)
forces smaller (larger) than $F_M$ ($F_m$). In particular, for $F_{p}$
close enough 
to $F_{M}$, the folded configuration becomes
thermodynamically metastable (it corresponds to a local
minimum while the unfolded configuration corresponds to the absolute
minimum thereof). The same is true (with the roles reversed) for
$F_{f}$ close 
to $F_{m}$: the unfolded configuration becomes
metastable (local minimum) and the folded configuration becomes stable
(absolute minimum). The escape time from the metastable states is
finite at finite temperature and it becomes infinite only in the zero
temperature limit (deterministic case). As the energy barrier between
the unfolded and folded configuration vanishes for $F_{p}\to
F_{M}^{-}$ ($F_{f}\to F_{m}^{+}$), these escape times are expected to
become smaller the closer the peak (final) force is to $F_{M}$
($F_{m}$). Note that Kramers rate theory requires large
  energy barriers (in units of $k_BT$) for separation of time scales
corresponding to intra- and interwell dynamics 
  \cite{HTyB90}. This condition no longer holds for $F$ very close to
$F_M$ and $F_m$.

Similarly to the previous discussion, the time intervals spent in the
virtual states become considerably longer than the deterministic
times, as they are proportional to the exponential of the barrier
energy in units of $k_BT$.  The stepwise deterministic unfolding of
Fig.~\ref{fig4}(a) is also affected strongly by the noise, which may
shorten or enlarge greatly the step duration, an effect observed in
experiments \cite{gar07}.  Throughout this section, we consider a
nondimensional temperature $\theta=k_BT/([F] L_c)=0.0024$
(corresponding to $T=300$K), independent of the damping constant. The
latter only selects the time unit $[t]=\gamma L_c/[F]$.

\begin{figure}[htbp]
\begin{center}
\includegraphics[width=3.25in]{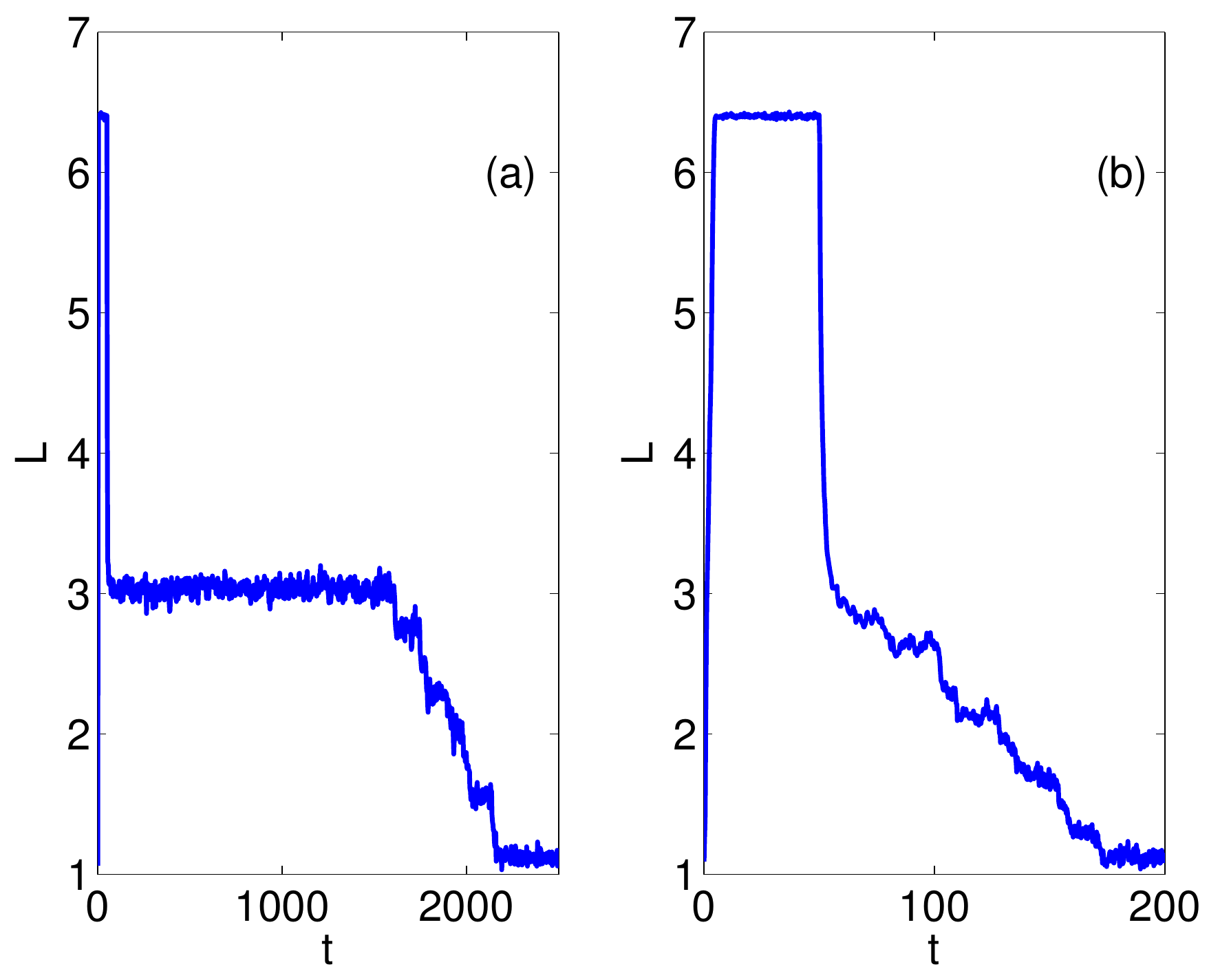}
\caption{Effect of noise on the length responses to force changes with
$F_p=1$ (100 pN) and (a) $F_f=0.117$ (11.7 pN) and (b) $F_f=0.11$ (11
pN). The nondimensional noise strength is $\theta=k_B
T/([F]L_c)=0.0024$. Note the different global timescales in
both graphs. When the force is suddenly decreased from $F_{p}$ to
$F_{f}$ ($t=50$), the system almost instantaneously falls on the
unfolded state corresponding to the final force $F_{f}$ in both
graphs. Afterwards, the jump to the corresponding completely folded
state at $F_{f}$ occurs over a much longer timescale for $F_{f}=0.117$,
which is further from $F_{m}$.}
\label{fig6}
\end{center}
\end{figure}

Firstly, let us see how thermal noise affects the refolding
stage. Fig.~\ref{fig6}(a) and (b) illustrate this by
depicting the length response to a force change where $F_p=1$ (100 pN,
just below $F_M=1.04$) and the final forces $F_{f}$ are
0.117 and $0.11$ (11.7 and 11pN), respectively,
both above but close to $F_{m}=0.104$. For a given temperature, the
more $F_{f}$ differs from $F_{m}$, the higher the barrier between the
unfolded and the folded configurations and the longer the timescale
for the refolding process. This is clearly shown in
Fig.~\ref{fig6}(a), in which the system refolds on a time scale that
is quite longer than the one in Fig.~\ref{fig6}(b). Moreover, the
length response after the force drops from $F_p$ to its final value is
quite similar in the cases of a polyprotein with 8 modules and of a
single module protein (not shown): The time scale for refolding is the
same in both cases, although not all the polyprotein modules refold
strictly at the same time in the case of Fig.~\ref{fig6}.
\begin{figure}
  \centering
\includegraphics[width=3.25in]{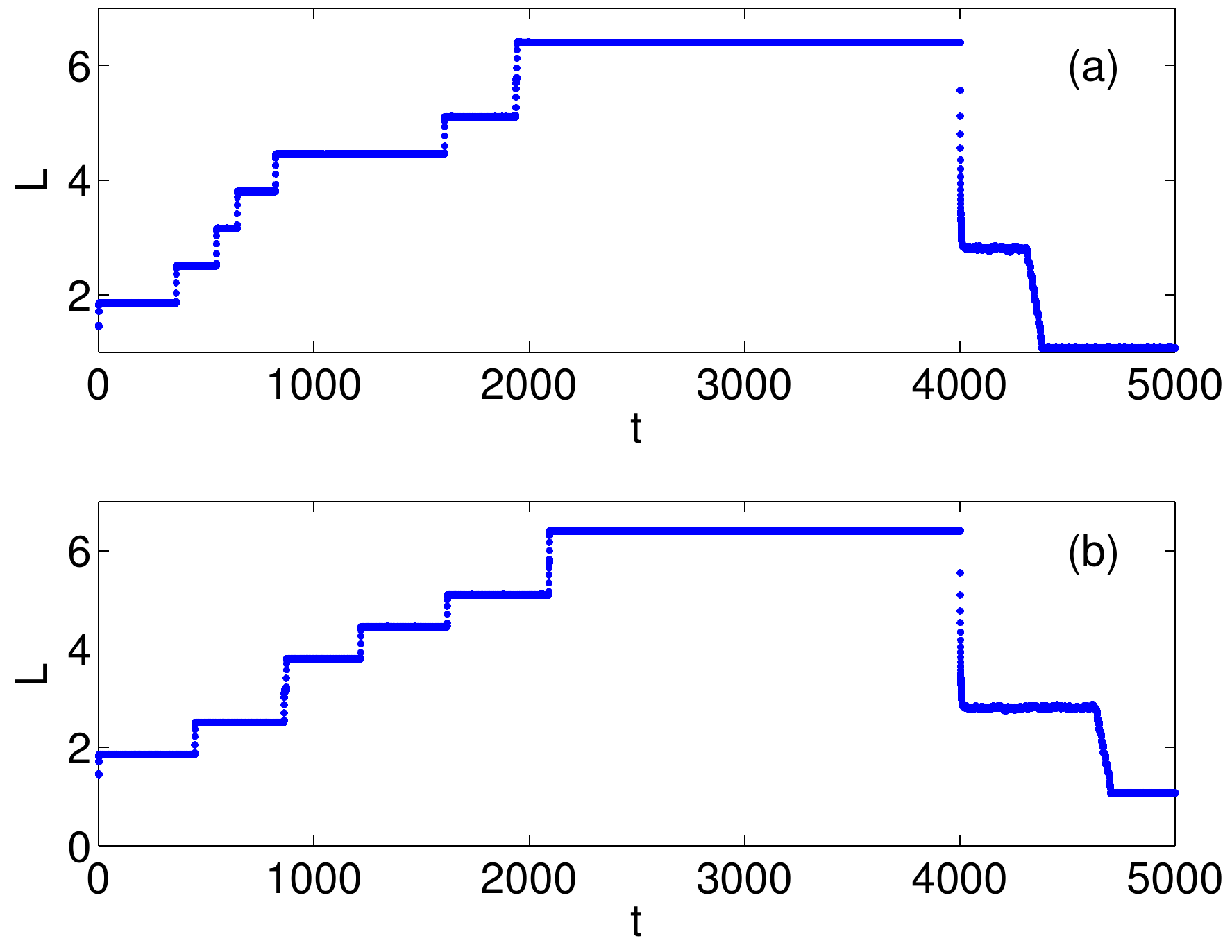}
  \caption{\label{newfig} 
Two realizations of the length responses to force changes with
nondimensional values $F_{in}=0.2$ (initial), $F_p= 1$ (peak),
$F_f=0.108$ (final). Note (a) the much longer timescale for the
unfolding process, as compared to the deterministic case shown in
Fig.~\ref{fig4}, and (b) the nonuniform duration of the stepwise
unfolding process. The final length is the same in both
cases.}
\end{figure}

Let us consider now how noise affects the unfolding stage. In
Fig.~\ref{newfig}, we show two realizations of the length response to
a force protocol like the one in Fig.~\ref{fig4}(b), with
initial, peak and final forces
$F_{in}=0.2>F_{m}$, $F_{p}=1$, and
$F_{f}=0.108>F_{m}$, respectively. The observed
stepwise unfolding is strongly affected by noise: (a) the system
evolves on a time scale roughly 100 times larger than that of
Fig.~\ref{fig4}, and (b) the steps are no longer uniform. The finite
value of the temperature may greatly shorten or enlarge the step
duration, an effect observed in experiments \cite{gar07}. The
refolding stage is similar to that in Fig.~\ref{fig6}. When
the force is decreased to $F_{f}$ at $t=4000$, all the modules remain
unfolded but their length almost instantaneously decrease to the value
$u^{(3)}(F_{f})$. Then, after spending a stochastic time (longer in
(b) than in (a)) in this metastable state, all the modules finally
refold to the thermodynamic stable state $u^{(1)}(F_{f})$. Similar to
Fig.~\ref{fig6}, the modules do not refold strictly at the same time
but the details are not appreciable due to the longer time span in
Fig.~\ref{newfig}.

\section{Discussion}

In force-clamp experiments exhibiting stepwise unfolding, polyproteins
are operating near critical conditions corresponding to the
non-equilibrium depinning transition for wave fronts
\cite{car03}. This may seem reminiscent of the idea that
muscle materials are finely tuned to operate close to a mean-field
equilibrium phase transition \cite{car13}. However, including
mean-field couplings as in \cite{car13} dramatically changes the
dynamics by precluding stepwise unfolding: The \textit{short-range}
couplings between modules are at the root of the depinning transition
that causes the poorly understood stepwise unfolding observed in
experiments for certain values of the peak force. 
  
Depinning of wave fronts is a general phenomenon in spatially discrete
systems \cite{car01,car03,TyV10} and it is behind observed
behavior in systems quite removed from proteins. For instance,
nonlinear charge transport in weakly coupled dc voltage
  biased semiconductor SLs produces current--voltage
curves alike FECs in length-clamp experiments (current is
assimilated to force and voltage to extension) \cite{BGr05}. Stepwise
unfolding of the force-clamp protein could have a counterpart in dc
current biased SLs but the necessary experiments do not yet exist, as
there are no precise current controlled experiments on these
nanostructures.

We have provided a unified framework involving quasi-stationary
virtual states to understand quite different observed behaviors, such
as simultaneous vs.~stepwise unfolding or three-stage vs.~two-stage
refolding. These virtual states are the closest stable configurations
that the modules can attain when the forces are outside (but close to)
the metastability region, that is, the region of forces in which the
unfolded and folded configurations coexist.

We also predict novel behaviors, such as the sequential character of
stepwise unfolding. Thus our work opens new attainable perspectives in
the experimental investigation of tethered biomolecules. In real
experiments, unfolding may not be sequential if the heterogeneity in
the potential is larger than the intermodule spring potential
energy. Then additional simulations of our model would predict the
unfolding order.

\section{Appendix: Nondimensional model}
\label{sec:0}
We measure force, extensions and time in the units: $[F]=100$ pN,
$L_c=30$ nm and $[t]= \gamma L_c/[F]=k_BTL_c/(D[F])$,
respectively. The equations of the model are
\begin{eqnarray}
 \dot{u}_j &=& F - V'(u_j) + \kappa\, (u_{j+1}+u_{j-1}-2u_j)\nonumber\\
 && + \sqrt{2\theta}\, \xi_j(t),  \label{s1}\\
V(u)&=& \mu\!\left\{\left[1-e^{-\beta(u-\rho)}\right]^2-1 \right. \nonumber \\
&&
\left.\quad  +A\!\left(\frac{1}{1-u}-1-u+2u^2\right) \right\},    \label{s2}
\end{eqnarray}
where $\mu=U_0/(L_c[F])$, $\beta=2bL_c/R_c$, $\rho=R_c/L_c$,
$\kappa=kL_c/[F]$, $A=k_B TL_c/(4PU_0) $, $\theta=k_B T/([F]\,
L_c)=0.0014$ and the $\xi_j(t)$ are i.i.d.\ zero-mean delta-correlated
white noises. Note that $\theta$ is independent of the diffusion
constant $D$, which sets the unit of time $[t]$.

\subsection{Stepwise unfolding}
To explain stepwise unfolding when $\theta=0$ and $F$ has increased
abruptly to $F_p\in(F_{c2},F_M)$ from $F_{in}$, assume that one module
has stretched to $u^{(3)}(F_{c2})$ and the others to
$u^{(1)}(F_{c2})$ for a critical force $F_{c2}$ (slightly below
$F_M$) such that 
$\kappa[u^{(1)}(F_{c2})+u^{(3)}(F_{c2})]=V'(u^*)+2\kappa u^*$ and 
$2\kappa+V''(u^*)=0$,
for $u_J=u^*$. We have a wave front joining a
domain with $N-1$ modules of extension $u^{(1)}(F_{c2})$ and one
unfolded module of extension $u^{(3)}(F_{c2})$. Let $u_J$ be the
extension of the module adjacent to the unfolded one. In (\ref{s1}) we
have $u_j=u^{(1)}(F_{c2})$ for $j<J$ and $u_{j}=u^{(3)}(F_{c2})$ for
$j>J$. Then expanding the right hand side of (\ref{s1}) in powers of
$(u_J-u^*)$, we obtain 
\begin{eqnarray}
&&\dot{u}_J \sim F_p-F_{c2}-\frac{1}{2}V'''(u^*)(u_J-u^*)^2,
\label{s3}
\end{eqnarray}
provided that
  $u^{(1)}(F_{c2})+u^{(3)}(F_{c2})=2u^*+\frac{V'(u^*)}{\kappa}$ and
  $V''(u^*)+2\kappa=0$.  Since $\kappa$ is small, $u^*$ is close to
$u_M$ and $F_{c2}<F_p<F_M$ are close. Then $V'''(u^*)\approx
V'''(u_M)<0$ and (\ref{s3}) has the solution $u_J =
  u^*+\frac{2\Gamma}{|V'''(u^*)|} \, \tan\!\left[\Gamma
    (t-t_J)\right]$,  where $t_J$ is a constant and
$\Gamma=\sqrt{(F_p-F_{c2})|V'''(u^*)|/2}$. Notice that $u_J(t_J)=u^*$
and that the tangent function becomes $\pm\infty$ when
$\Gamma (t-t_J)=\pm\pi/2$. After the argument of the
tangent function reaches $\pi/2$, at $t=t_J+\pi/(2\Gamma)$, $u_J$
jumps to $u^{(3)}(F_{c2})$, and the point $u_{J-1}(t)$ becomes
active. This means that the $J$th module has unfolded, the wave front
has advanced one step to the left and $u_{J-1}(t)$ satisfies
(\ref{s3}) for $|t-t_{J-1}|<\pi/(2\Gamma)$, where
$t_{J-1}=t_J+\pi/(2\Gamma)$. The duration of the steps between jumps
of the wave front is $(t_{j-1}-t_J)=\pi/(2\Gamma)$. Details of the
jumps and the matching between jumps are given in \cite{car01,car03}
for the saltatory motion of wave fronts near the depinning transition.

\subsection{Simultaneous unfolding and refolding}
Simultaneous module unfolding and refolding imply evolution to virtual
states at $u_j=u_M$ and $u_j=u_m$ (for all $j$),
respectively. Following a line of reasoning similar to that
in the last paragraph, we find that the $u_j$ are near
$u^*$ during a long time
$\pi/\sqrt{2|F-F^*|\,|V'''(u^*)|}$ ($F^*$ is $F_M$ or
  $F_m$).

\acknowledgments
This work has been supported by the Spanish Ministerio de Econom\'\i a y Competitividad grants FIS2011-28838-C02-01 (LLB),  FIS2011-28838-C02-02 (AC), and FIS2011-24460 (AP).


\begin{thebibliography}{28}
\bibitem{alb98}
 \Name{Alberts, B.}  \Review{Cell}  \Vol{92} \Year{1998} \Page{291}.

\bibitem{obe08}
\Name{Oberhauser, A.F. \and Carri\'on-V\'azquez, M.}  \Review{J. Biol. Chem.} \Vol{283} \Year{2008} \Page{6617}.

\bibitem{lin08}
\Name{Linke, W. A.} 
\Review{Cardiovasc. Res.} \Vol{77} \Year{2008} \Page{637}.

\bibitem{thirumalai13} \Name{Gruebele, M. \and D. Thirumalai, D.} \Review{J. Chem. Phys.} \Vol{139} \Year{2013} \Page{121701}.

\bibitem{car99} \Name{Carrion-V\'azquez, M. \etal } \Review{Proc. Natl. Acad. Sci. USA} \Vol{ 96} \Year{1999} \Page{3694}.

\bibitem{fis00} \Name{Fisher, T. E. \etal } 
\Review{Nature Struct. Biol.} \Vol{7} \Year{2000} \Page{719}.

\bibitem{MyD12} \Name{Marszalek. P. E. \and Dufr\^ene, Y. F.} 
\Review{Chem. Soc. Rev.} \Vol{41} \Year{2012} \Page{3523}.

\bibitem{rit06jpcm}
\Name{Ritort, F.} 
\Review{J. Phys: Condens. Matter} \Vol{18} \Year{2006} \Page{R531}.

\bibitem{lip01} \Name{Liphardt, J. \etal } \Review{Science} \Vol{292} \Year{2001} \Page{733}.

\bibitem{bus03} \Name{Bustamante, C. \etal } 
\Review{Nature} \Vol{ 421} \Year{2003} \Page{423}.

\bibitem{lip02} \Name{Liphardt, J. \etal } 
\Review{Science} \Vol{296} \Year{2002} \Page{1832}.

\bibitem{PCyB13} \Name{Prados, A. \etal } \Review{Phys. Rev. E} \Vol{88} \Year{2013} \Page{012704}.

\bibitem{smi96} \Name{Smith, S. B. \etal } \Review{Science} \Vol{271} \Year{1996} \Page{795}.

\bibitem{hug10} \Name{Huguet, J. M.} 
Ph.D. Thesis, Univ\textrm{.} Barcelona \Year{2010}.

\bibitem{cao08}
\Name{ Cao, Y. \etal } 
\Review{Biophys. J.} \Vol{95} \Year{2008} \Page{782}.

\bibitem{PCyB12} \Name{Prados, A. \etal } \Review{Phys Rev E} \Vol{85} \Year{2012} \Page{031125}.

\bibitem{li04} \Name{Fernandez, J. M. \and Li, H.} \Review{Science} \Vol{303} \Year{2004} \Page{1674}.

\bibitem{ber10} \Name{Berkovich, R. \etal } \Review{Biophys. J.} \Vol{98} \Year{2010} \Page{2692}.

\bibitem{wal07} \Name{Walther, K. A. \etal } \Review{Proc. Natl. Acad. Sci. USA} \Vol{104} \Year{2007} \Page{7916}.

\bibitem{BGr05}
\Name{Bonilla, L. L., \and Grahn, H. T.} \Review{Rep. Prog. Phys.} \Vol{68} \Year{2005} \Page{577}.

\bibitem{lan12} \Name{Lannon, H. \etal } \Review{Phys Rev Lett} \Vol{110} \Year{2013} \Page{128301}.

\bibitem{ByG11} \Name{Benichou, I. \and Givli, S.} \Review{Appl. Phys. Lett.} \Vol{98} \Year{2011} \Page{091904}. 


\bibitem{PyT02} \Name{Puglisi, G. \and Truskinovsky, L.} 
\Review{J. Mech. Phys. Solids} \Vol{50} \Year{2002} \Page{165}.


\bibitem{tommasi} \Name{De Tommasi, D. \etal } \Review{J. R. Soc. Interface} \Vol{10} \Year{2013} \Page{20130651}.

\bibitem{lee10} \Name{Lee, W. \etal } 
\Review{J. Biol. Chem.} \Vol{285} \Year{2010} \Page{38167}.

\bibitem{gar07} \Name{Garcia-Manyes, S. \etal } \Review{Biophys. J.} \Vol{93} \Year{2007} \Page{2436}.

\bibitem{car01} \Name{Carpio, A. \and Bonilla, L. L.} \Review{Phys. Rev. Lett.} \Vol{86} \Year{2001} \Page{6034}. 

\bibitem{car03}\Name{Carpio, A. \and Bonilla, L. L.}  \Review{SIAM J. Appl. Math.} \Vol{63} \Year{2003} \Page{1056}. 

\bibitem{TyV10} \Name{Truskinovsky, L. \and Vainchtein, A.} 
\Review{Continuum Mech. Thermodyn.} \Vol{22} \Year{2010} \Page{485}.

\bibitem{fis98} \Name{Fisher, D. S.} \Review{Phys. Rep.} \Vol{301} \Year{1998} \Page{113}. 

\bibitem{HTyB90} \Name{Hanggi, P. \etal }  \Review{Rev.~Mod.~Phys.} \Vol{62} \Year{1990} \Page{251}.

\bibitem{car13} \Name{Caruel, M. \etal } 
\Review{Phys. Rev. Lett.} \Vol{110} \Year{2013} \Page{248103}.

\end{thebibliography}
\end{document}